\documentclass[fleqn,10pt,inline]{wlscirep}
\usepackage[utf8]{inputenc}
\usepackage[T1]{fontenc}

\title{Ruhr Hand Motion Catalog of Human Center-Out Transport Trajectories in 3D Task-Space Captured by a Redundant Measurement System \\[1ex] \LARGE \textit{Data Descriptor}}

\author[1,2,3,+]{Tim Sziburis}
\author[2,4,+]{Susanne Blex}
\author[1]{Tobias Glasmachers}
\author[2]{Ioannis Iossifidis}
\affil[1]{Institute for Neural Computation, Ruhr University Bochum, 44801 Bochum, Germany}
\affil[2]{Institute of Computer Science, Ruhr West University of Applied Sciences, 45479 Mülheim an der Ruhr, Germany}
\affil[3]{Faculty of Electrical Engineering and Information Technology, Ruhr University Bochum, 44801 Bochum, Germany}
\affil[4]{Astronomical Institute, Ruhr University Bochum, 44801 Bochum, Germany}
\affil[+]{these authors contributed equally to this work}

\begin{abstract}
Neurological conditions are a major source of movement disorders. Motion modelling and variability analysis have the potential to identify pathology but require profound data. We introduce a systematic dataset of 3D center-out task-space trajectories of human hand transport movements in a natural setting. The transport tasks of this study consist of grasping a cylindric object from a unified start position and transporting it to one of nine target locations in unconstrained operational space. The measurement procedure is automatized to record ten trials per target location. With that, the dataset consists of 90 movement trajectories for each hand of 31 participants without known movement disorders. The participants are aged between 21 and 78 years, covering a wide range. Data are recorded redundantly by both an optical tracking system and an IMU sensor. As opposed to the stationary capturing system, the IMU can be considered as a portable, low-cost and energy-efficient alternative to be implemented on embedded systems.
\end{abstract}
\begin{document}

\flushbottom
\maketitle
\thispagestyle{empty}

\section*{Background \& Summary}

The experimental observation and analysis of reaching tasks is an elementary means to develop a systematic understanding of movement generation. As a specific form of a reaching task (also referred to as aiming task), a transport (or transportation) task can be understood as the combination of a first reach-to-grasp movement, the grasping of the object itself, followed by transporting the object and reaching towards a target, before positioning it at the target location. In this study, we measured the movement trajectories during the time of transporting the grasped object, i.\,e. without consideration of grasping activity.

Over the last decades, there have been various studies in the discipline of neuroscience comprising different variants and phases of reaching task experiments. Studies with human participants include analyses of kinematic coordination and muscle activity, such as with regard to trial consistency and coupling between distal and proximal kinematic chain trajectories\cite{lacquaniti_coordination_1982}, or with specific focus on three-dimensional target positioning\cite{vandenberghe_three-dimensional_2010}. Furthermore, there exist studies about hand selection strategy \cite{stins_kinematic_2001}, decision-making\cite{chapman_reaching_2010}, and motor sequence learning \cite{moisello_serial_2009}. Reaching experiments were also conducted with non-human primates: Georgopolous et al. described the activation of specific direction-coded neural populations during two-dimensional\cite{georgopoulos_relations_1982} and three-dimensional\cite{georgopoulos_neuronal_1986,georgopoulos_primate_1988} reaching tasks.

Nevertheless, datasets of the experiments are sparsely publicly available, do not specifically target object transportation, or inquire 2D movements with touchscreens or 3D movements in virtual reality.

A collection of published datasets is gathered in the Database for Reaching Experiments and Models (DREAM) from Northwestern University\cite{walker_database_2013}. In this database, several datasets from reaching tasks are published in combination with specific tools and models, also with special focus on neural activity\cite{dream}. Conducted with different modalities of recording, it comprises datasets of reaching studies on:
\begin{itemize}[noitemsep]
    \item observations of reaching tasks without primary motor cortex population vectors pointing in hand motion direction, measuring neural activity in monkeys\cite{scott_dissociation_2001},
    \item probabilistic models during sensorimotor learning via optical finger tracking in the plane \cite{kording_bayesian_2004},
    \item neural tuning of movement space in environments of differing complexity, measured by grasping a robotic arm exerting perturbing forces to change movement direction\cite{thoroughman_rapid_2005},
    \item generalization of dynamics learning to novel directions by interpolation, horizontally measured by optical encoders of a robotic handle also exerting forces \cite{mattar_modifiability_2007},
    \item movement timing and speed–accuracy trade-off (Fitt's law), measured with a stylus on a tablet\cite{young_target-directed_2009},
    \item nervous system motor adaptation to errors, proposing a non-linear model, recorded by a robotic manipulandum\cite{wei_relevance_2009},
    \item influences of perturbations, measured by moving a horizontally restricted robot arm\cite{wei_nervous_2010},
    \item force production and generalization with differing movement amplitudes while moving a robotic manipulandum\cite{mattar_generalization_2010},
    \item multi-sensory integration while moving a haptic device in a horizontal manner and also measuring electrooculography\cite{burns_multi-sensory_2010},
    \item effects of motor-learning on somatosensory plasticity, experiments conducted with a planar robotic arm handle with optical encoders and force-torque sensors for measurements\cite{ostry_somatosensory_2010}, 
    \item tuning curve stability of neural representation of limb motion during center-out reaching of monkeys, measured from primary motor cortex\cite{stevenson_statistical_2011},
    \item changes of neural functional connectivity after motor learning, measured by a planar robotic handle together with MRI scans \cite{vahdat_functionally_2011},
    \item neuroprosthetic control by recording a stylus attached to a robot in combination with eye- and head-tracking as well as electromyography \cite{corbett_real-time_2012},
    \item the effect of uncertainty on the generalization of visuomotor rotations by optical tracking a finger used for two-dimensional cursor control \cite{fernandes_generalization_2012}, and
    \item motion target and trajectory decoding of center-out and random target tasks from neural recordings of monkeys, moving and tracking a two-link manipulandum \cite{flint_accurate_2012}.
\end{itemize}
However, none of these studies specifically examines the object transport after grasping, which is the target of this study. 

Other experiments involving data from actual transport tasks with healthy participants consisted of studies regarding force trajectories\cite{hejdukova_grip_2002}, correction motion in the presence of moving targets\cite{danion_can_2007}, or age-dependent motor coordination\cite{huntley_older_2017}. Several transport motion analyses with participants suffering from diseases captured data in comparison to a healthy control group. These included the examination of force control in Huntington's disease \cite{quinn_altered_2001}, movement dexterity of Parkinson's disease patients\cite{hejdukova_manual_2003}, or eye-hand coordination under hemiparetic cerebral palsy\cite{verrel_eyehand_2008}. However, these studies targeted specific pre-selected movement properties that were supposedly characteristic for the individual use case and lack the systematic variation of the set-up such as regarding target positioning. Therefore, the data from these studies, if available, would be lacking detailed and systematic insights into the general processes and methodology behind movement generation.

With specific regard to systematic hand transportation experiments with varying parameters, Grimme et al. (2012)\cite{grimme_naturalistic_2012} recorded natural human arm motion for two target directions of transporting an object over varying distances as well as obstacle heights and positions. While this first experiment was conducted with ten healthy participants, a second experiment with five participants (again, without known movement disorders) also included further obstacle configurations of one target direction. They analyzed specific invariants and properties of such motion, including isochrony, planarity and the decomposition of movements into a transport and a lift/descend primitive. Based on extended datasets gathered in further experiments\cite{Grimme2015}, they furthermore describe an obstacle-dependent primitive. A subset of that data was published\cite{bochum}.

Besides general public availability, our novel dataset provides systematicity with regard to target positioning and consists of methodologically captured hand transport movements in 3D. Furthermore, to our knowledge, it is the first dataset comprising transport trajectory recordings from both hands with target-randomized trial settings. The center-out setting of the task can provide information on direction-dependent motion components and factors. The general set-up makes it relevant for movement modelling as well as for comparison with later studies on movement impairments. 

None of the aforementioned studies and data utilized more than one system for parallel measurement, since optical systems are considered as precise state-of-the-art techniques for motion capture. 
However, they are usually not suitable for embedded applications. Instead, portable sensors such as inertial measurement units (IMUs) might be favorable, for example in medical diagnosis scenarios requiring flexible applicability and mobility.

Thus, our presented dataset provides synchronized capturing from both systems to evaluate the feasibility of the portable system.
\section*{Methods}

\subsection*{Participation} \label{methods_participation}
The experiments were conducted in November and December 2022 in the eHealth laboratory of the Ruhr West University of Applied Sciences. 
Experimental data was recorded from thirty-one participants without known movement disorders. After getting introduced to the experimental procedure and providing informed consent, the participants filled a questionnaire with the following basic information: \begin{itemize*}[noitemsep]
    \item age,
    \item gender,
    \item body height,
    \item general physical condition, and
    \item former experience with motion capture experiments.
\end{itemize*}

Furthermore, they conducted the 10-item Edinburgh Handedness Inventory\cite{oldfield_assessment_1971} in order to calculate their handedness score. The following aspects were addressed: \begin{itemize*}[noitemsep]
    \item writing,
    \item painting,
    \item throwing,
    \item scissors,
    \item tooth brush,
    \item knife (without fork),
    \item spoon,
    \item broom (upper hand),
    \item lighting a match,
    \item opening the lid of a box,
    \item preferred foot, and
    \item preferred eye.
\end{itemize*}

\begin{figure}
    \begin{center}
     \includegraphics[width=0.49\textwidth]{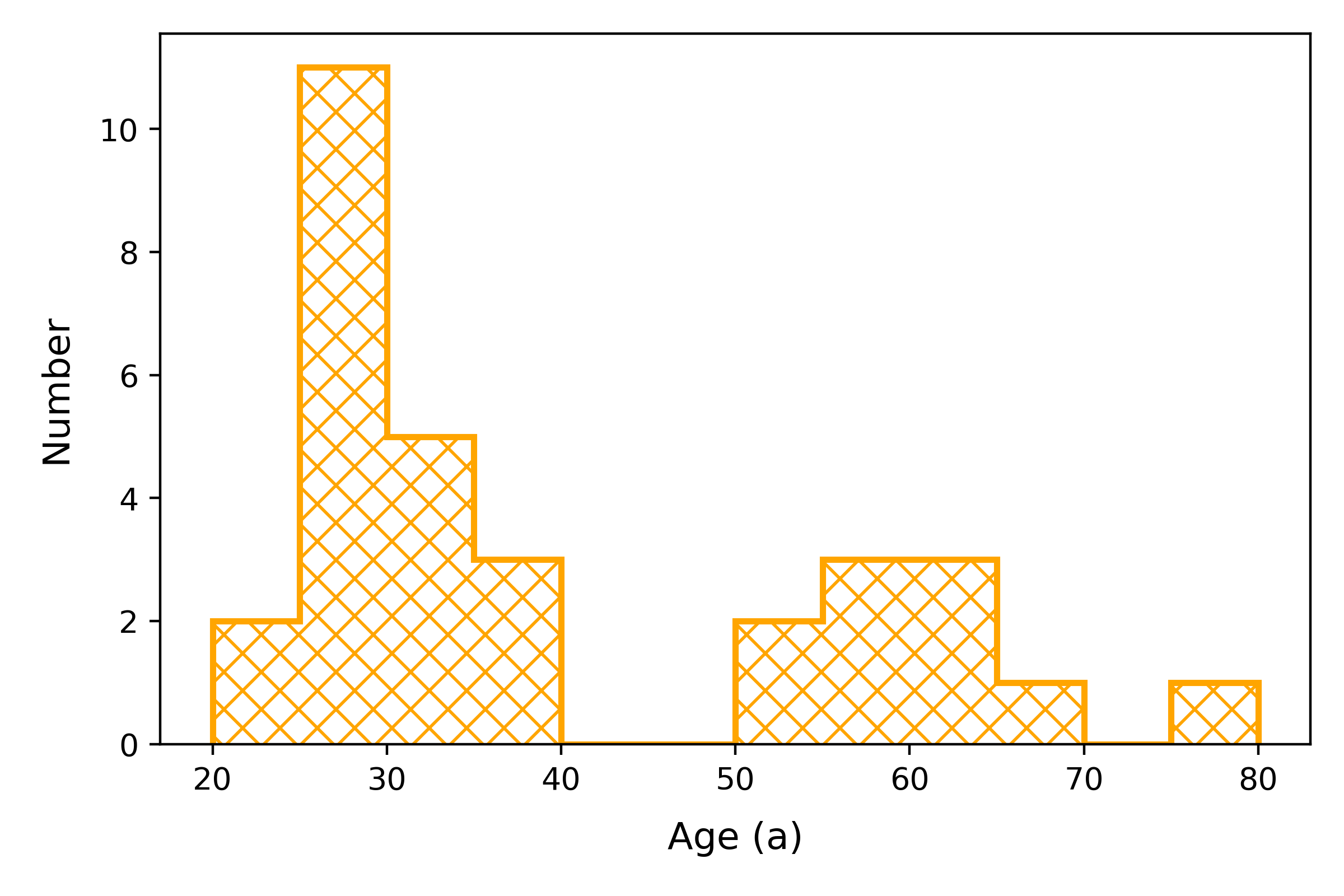}
     \hfill\includegraphics[width=0.49\textwidth]{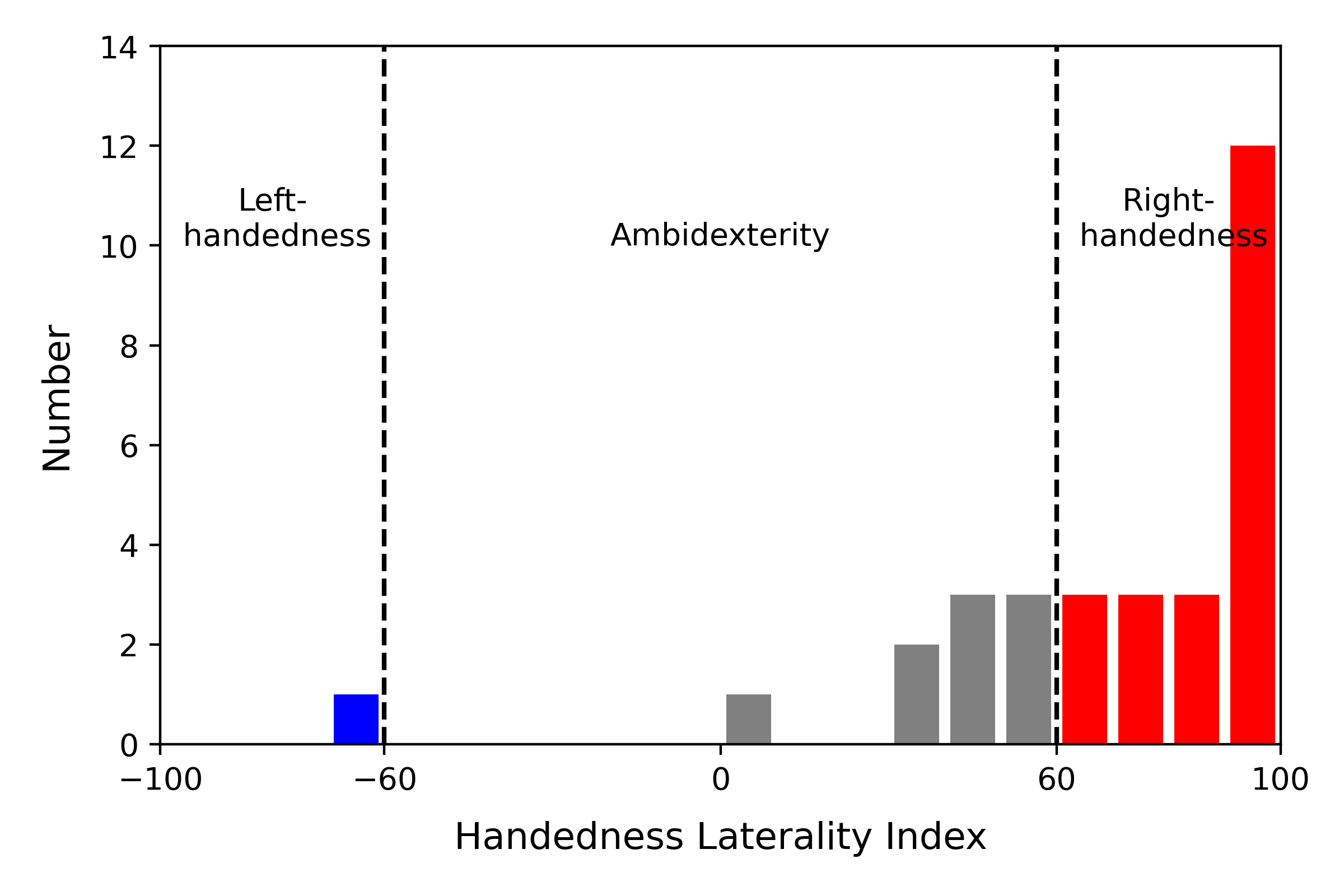}
     
     \caption[Overview of age and handedness of the study participants]{Overview of age and handedness of the study participants.
     \textit{Left:} Histogram of the study participants ages. \textit{Right:} Histogram of EHI handedness laterality indices of the study participants. Marked in color are the assignments to right-handed, left-handed, and ambidextrous individuals.}
     \label{fig:hists}
    \end{center}
   \end{figure}

The participants, 11 female and 20 male, were in a range of age between 21 and 78. \autoref{fig:hists} shows histograms of age and handedness of the study participants.

\subsection*{Experimental Design}
\begin{figure}[ht]
\centering
\includegraphics[width=0.3\linewidth]{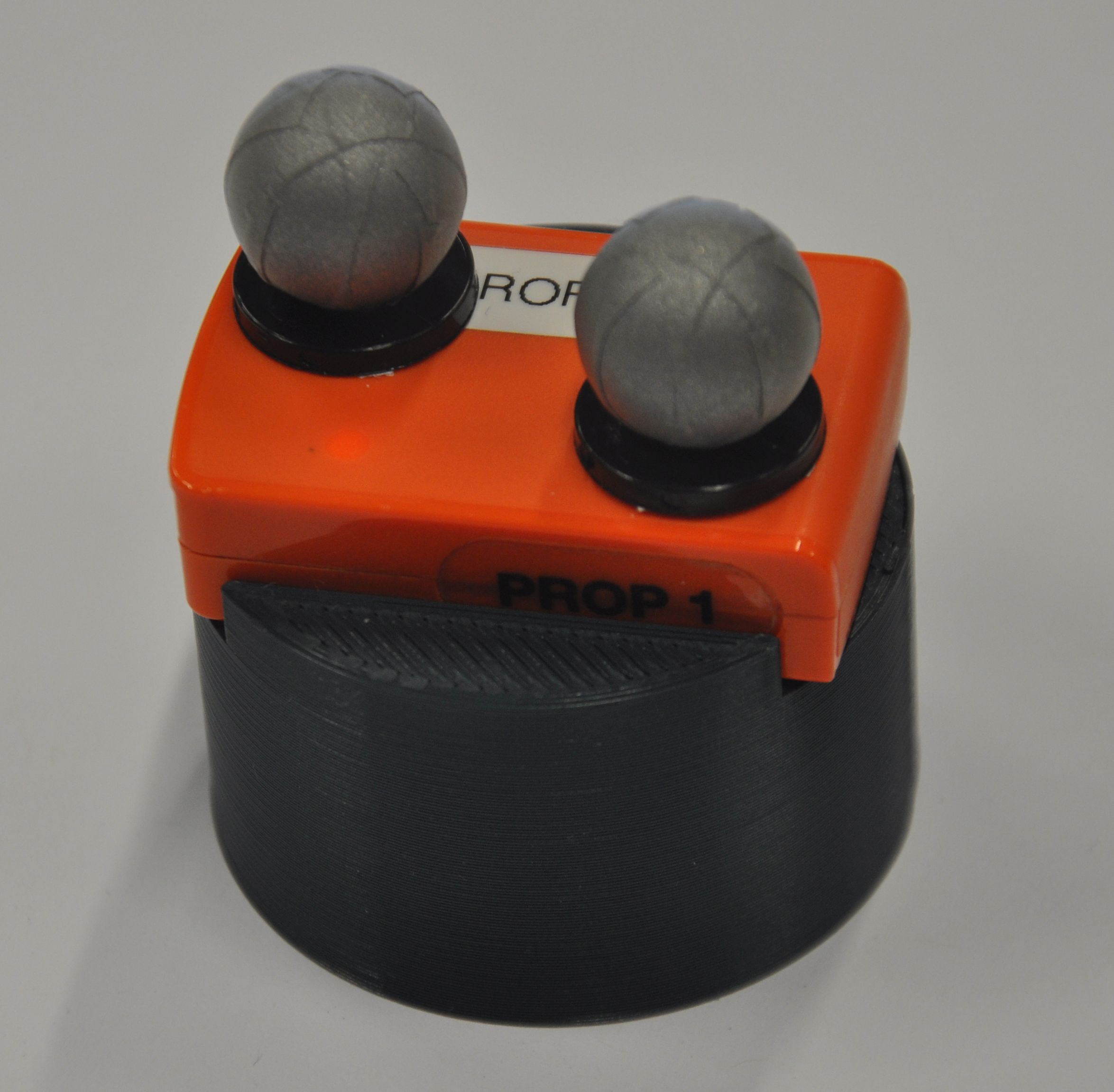}
\caption{Cylindric transport object with IMU sensor on the top, fixed by latches. Furthermore attached on the top, two reflective markers for the optical motion capture system.}
\label{fig:prop}
\end{figure}

Transport trajectories were measured by means of two motion capture systems in parallel, while participants moved a cylindrical solid body from a start position to one of nine target positions. The cylindrical object had a diameter of 5\,cm and a height of 2.5\,cm. Being individually 3D-printed, the cylinder provided the possibility to lock the used sensor object by latches, see \autoref{fig:prop}.

In this way, a single inertial measurement unit (IMU) sensor was attached to the cylindrical transport object and recorded movement data. Additionally, six cameras were positioned in the room which recorded two reflection elements.

These reflective spheres were glued at the top of the IMU sensor diagonally so that at least one of them was always visible from any perspective angle.
Both systems were time-synchronized via NTP. The cameras were positioned in the room in a way that occlusions caused by the participants' movements were avoided. Furthermore, the positions of the cameras, their aperture, focus, and zoom settings were adjusted so that the tracking object was recognizable from all cameras and that no reflection disturbances from other sources appeared. For this reason, the windows of the laboratory were shaded, and ceiling lighting was switched on to guarantee identical illumination conditions for all participants.

The IMU data communication base station was positioned at one corner of the table and aligned with its edges, not disturbing or distracting the participants.
The experimental location in the laboratory room was selected in a way that magnetic field perturbations influencing the IMU sensor were minimized. 

\begin{figure}[ht]
\centering
\includegraphics[width=0.8\linewidth]{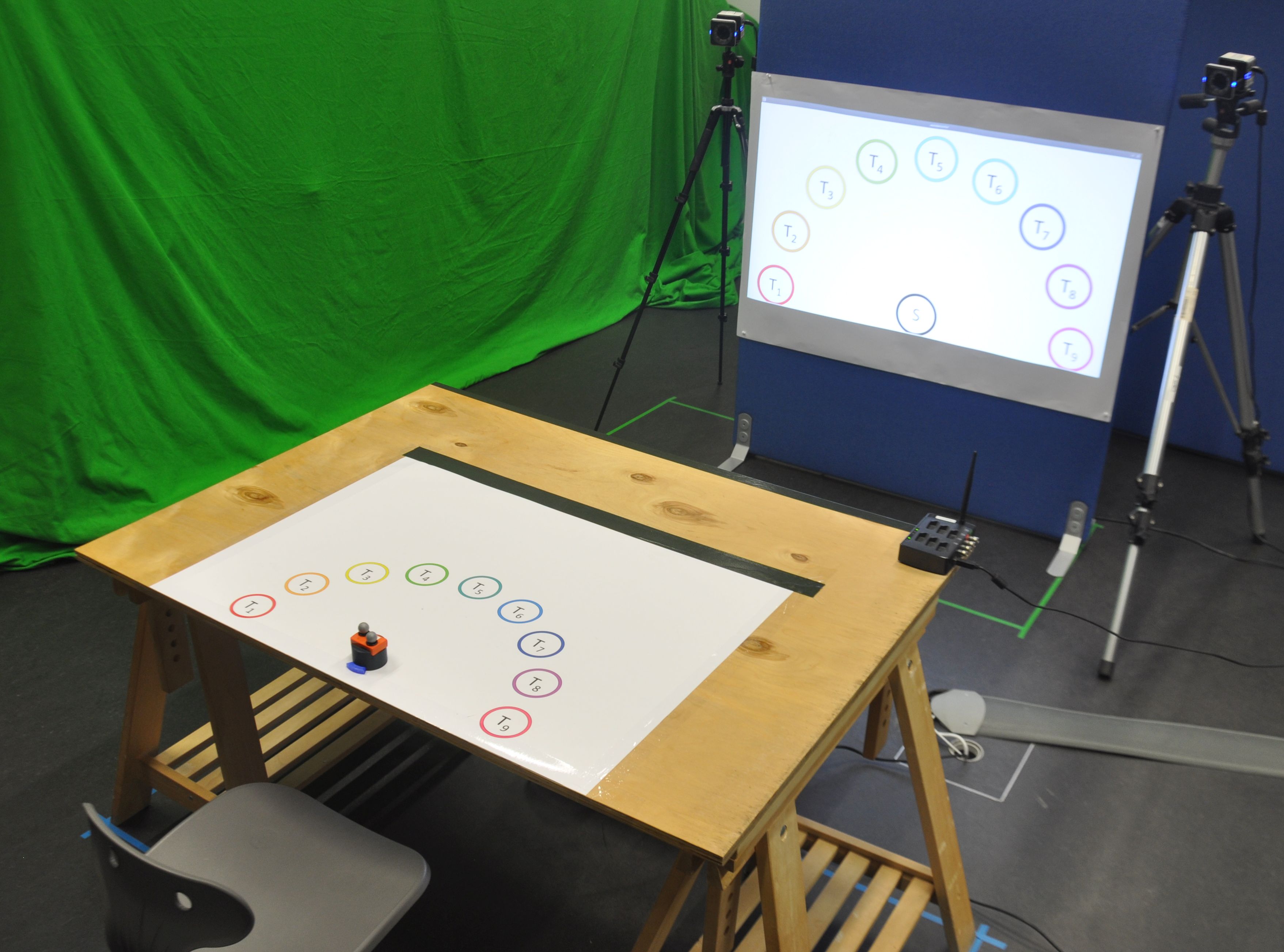}
\caption{Experimental set-up. While seating, participants were asked to perform 3D transport tasks from the start to randomized target locations after an acoustic start signal, visually announced in the projection.}
\label{fig:setup}
\end{figure}

Participants were seated on a chair in front of a table, which were both adapted to their individual body dimensions (see also section Data Acquisition Workflow). The table was made out of wood to further reduce electromagnetic influences. Chair and table positions were fixed in the room, adjustments were possible via height and tabletop position, respectively. On the table, the start position and the target positions were marked by circles, see \autoref{fig:setup}. The targets were equidistantly positioned on a semicircle, with the start position in the center and a radius of 25\,cm. The start and target circles themselves had a radius of 6\,cm. The center of the start circle was positioned at 6\,cm from the edge of the table. A 3D-printed docking block in the form of a short segment of a circle was glued to the printed starting circle (blue block in \autoref{fig:setup}). With that, an identical start position could be guaranteed for all trials.

For displaying target cues, a projector was positioned at the bottom and in front of the table, not observable by the participants so that they experienced no distraction. The projection was displayed on a canvas further in front and directly visible. Auditory start and stop signals were played via a loudspeaker integrated in the projector.

To avoid rhythmic movement patterns and specific time dependencies such as anticipatory behavior\cite{tsunoda_anticipation_2011}, random delays were introduced between the visual target cue and the acoustic start signal.

Each participant performed two sessions, one for each hand. Half of the participants started with the left-hand session, the other with the right-hand session. Both sessions were performed one after another, separated by a small break of some minutes.

Combining the transport cylinder with the IMU sensor and the two reflective elements for optical tracking, the overall transport object had a weight of 41\,g in total.

\subsection*{Instrumentation}
\begin{figure}[ht]
\centering
\includegraphics[width=0.8\linewidth]{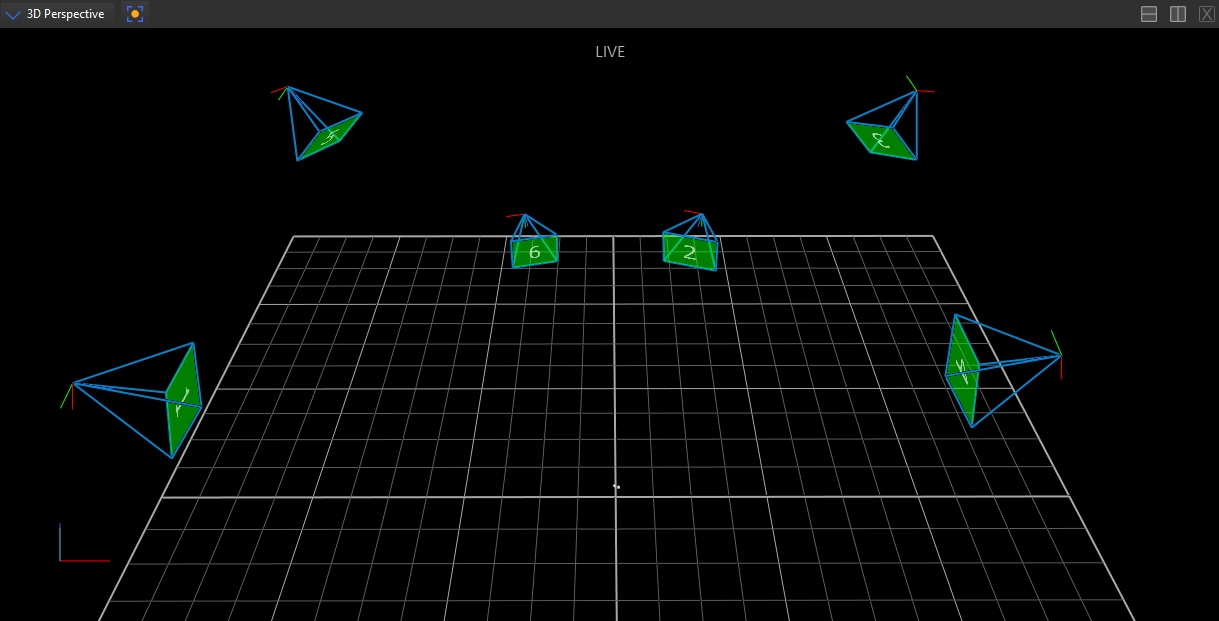}
\caption{Positioning of the Vicon Nexus optical motion capture cameras in the laboratory room, illustrated in the reconstructed world model. Moreover visible are the two reflection markers which are attached to the cylindric transport object, located at the start position.}
\label{fig:cameras}
\end{figure}

A state-of-the-art optical motion capture system (Vicon Nexus 2.14) comprised of six infrared cameras (Vicon Vero 1.3, with 1.3\,MP resolution, maximum frame rate 250\,Hz) was utilized to provide precise position reference data. These movements were measured by the continuous optical tracking of two reflection elements attached to the cylindric transport object. From the infrared recordings of the cameras that were able to detect the reflection elements in each frame, the individual three-dimensional positions were calculated in a world frame and combined to a course of positional data. The capturing rate was configured to the maximum possible frame rate of 250\,Hz. The camera positioning is visualized in \autoref{fig:cameras}.

The optical system with the cameras and the measurement computer were interconnected via RJ45 Ethernet ports. The measurement PC for the optical system was running a 64-bit Microsoft Windows 10 operating system with a 10 Gbit Ethernet controller.

The second measurement system, based on drift-corrected accelerometer data of a single state-of-the-art inertial measurement unit (IMU) from an Xsens MTw Awinda system, was used in parallel to evaluate the quality of a portable sensor set-up with embedded applicability. The following movement data were recorded by the IMU: accelerations (accelerometer), angular velocities (gyroscope), and orientations (magnetometer). 

For capturing the IMU data, the Xsens SDK from the Xsens MT Software Suite 2021.4 was utilized. Since an Xsens MTw system was used as opposed to the more recent MTi systems, the closed-source API with proprietary libraries had to be called instead of the open-source API. A C++ program was written which incorporated the communication with the Xsens Awinda v2 base station, addressing the single sensor (MTw2), as well as initiating the measurement and eventually recording the data. Both the base station and the sensor were running the most recent firmware (4.6.0).

The base station was connected via USB to a measurement laptop running a 64-bit Debian GNU/Linux 11 "Bullseye". In order to take advantage of real-time capabilities, a specific kernel, namely Liquorix 6.0 with preemptive scheduling was loaded. 

Based on an example for the utilization of the Xsens SDK, the recording program was extended by a QT graphical user interface to allow the configuration of communication channel, frame rate and connected sensor. Moreover, this was applied to configure the experimental control and finally conduct the experiments, including visualizations and acoustic signals. Besides the actual data recordings and their exact start and stop timestamps for all trials, this program calculated and stored the randomized order of targets to reach as well as the durations of all randomization time periods.

\subsection*{Data Acquisition Workflow}
The participants were introduced to the study and received information on the experimental procedure as well as the purpose of the measurements. After they had the possibility to request clarification and details, they were asked to sign an informed consent form. If everything was clear, they also filled the questionnaire described in section Participation.

First, the wooden table and chair were adjusted for each individual participant in a way that they were sitting in an upright position. Furthermore, it was made sure that there was an angle of 90° inside of the elbow while positioning their hand palms on target 1 (the outer left) and target 9 (the outer right). Each participant's configuration (chair height and tabletop position) was noted.

After that, a calibration of the Vicon Nexus optical motion capture system was initiated via the active wand calibration method according to the official instructions from the manufacturer, i.\,e moving the wand evenly in and over all directions of the experimental space until each camera reached a count of 2000 calibration frames. Until a self-set world error threshold could be reached, the procedure was repeated. This was followed by masking the infrared cameras via the Vicon Nexus software in order to eliminate any potentially remaining externally disturbing reflections out of the particular region of measurements.

After successful calibration, the origin of the coordinate frame was set for the Vicon system by means of positioning the calibration wand horizontally on the table with the origin in the center of the starting circle.

We made sure that the battery of the IMU sensor was charged to at least 90\% and then continued with magnetic field mapping via the corresponding software tool from the MT Software Suite 2021.4 (Magnetic Field Mapper). 
For this, the IMU sensor, attached to the cylindrical transport object, had to be uniformly rotated around any possible axes, until the criteria for 3D calibration stated in the Xsens manual were fulfilled (for details, see section Technical Validation).

Before seating, the participants were asked to remove all electronic devices and metallic objects from their pockets and from their clothing to minimize electromagnetic disturbances. The task space was systematically adjusted to their individual body dimensions. As soon as they were correctly seated, they had the possibility to start a prototypical test trial to familiarize themselves with the visual projection, the timing characteristics with randomization, and the start and stop sounds.

\begin{figure}[ht]
\centering
\includegraphics[width=0.8\linewidth]{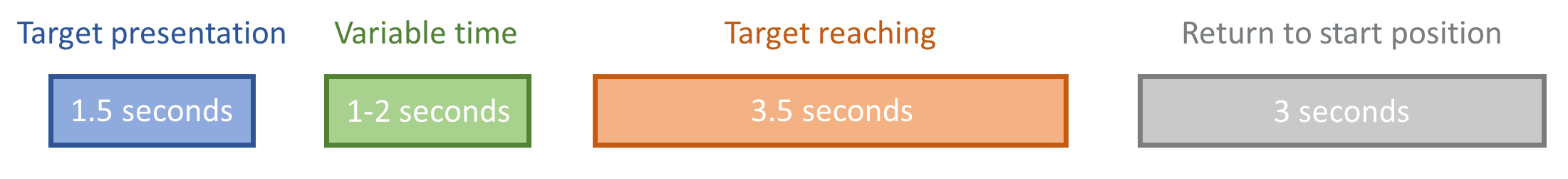}
\caption{Experimental workflow and measurement sequence of each trial with time periods.}
\label{fig:times}
\end{figure}

The recording procedure was automatized. The experimental sequence with the corresponding time spans per trial is depicted in \autoref{fig:times}.

Each participant performed two experimental sessions, one for each hand. In each transportation task, they grasped the object and moved it from the start to a target position on the labeled table. They were instructed to perform this movement as fast and as precise as possible. % as common in literature TODO

While the start position was always the same, the target position randomly varied between nine possibilities and was visually announced before an acoustic start signal appeared. The time between visual cue and acoustic start signal was randomized between one and two seconds. For each target, ten trials were asked to be performed for each hand. Conducted as a double-blind study, the appearing targets were neither known to the participants nor to the experimentators beforehand. After finishing a trial, participants had to stay still until an acoustic stop signal appeared, signalling to move the object back to the starting position again. Between the two sessions, a longer break was scheduled according to the individual needs. Furthermore, the participants could always introduce additional breaks between the trials if needed. However, nobody requested to make use of this possibility. One recording session usually took about 30 minutes (15 minutes per hand).

\subsection*{Data Processing}
While the IMU data capturing program recorded the data on a per-trial-basis and not during intermediate time periods, the Vicon system recorded the optical reflection data continuously from the start of the experimental session to its end. After finishing, this data stream was automatically cut into per-trial data via the stored IMU recording timestamps.

For both systems, a rotation was introduced to match the trajectory profiles. The trajectories were rotated in a way that the straight connection line between start and target maps onto the y-axis, calculated individually for each trial.

Since the time synchronization via NTP could have been subject to network delays, an additional fine alignment of time synchronization was introduced. This precise overall alignment for each trial could be realized via shifting the time courses to match the middle points of position on the y-axes.

Velocity and acceleration data of the optical system were calculated by differentiation, while for the IMU-based system velocity and position data were integrated from the acceleration after appropriate preprocessing. A fourth-order Butterworth low-pass filter was applied to all data to remove high-frequency noise from the trajectories (cut-off frequency 25\,Hz).

\subsubsection*{Optical Data}
The optical system utilized the infrared reflection recordings from all cameras which were able to perceive at least one marker in order to reconstruct the position course of the transported object in a world coordinate frame. For this, both markers had to be labeled in the Vicon Nexus software. If this assignment was not possible in a continuous manner, for example due to occlusions or the instantaneous appearance of further reflection signals, the labelling had to be performed for each self-contained sequence. If one optical marker was occluded by a participant's limb or head, the missing position of this marker in single frames could be padded by pattern fill: Since the two markers had a fixed position relative to each other, the pattern of movement could be inferred from the non-occluded marker. This was automatically manageable in the Vicon software. 

If in single frames both markers appeared to be not visible, also automatic spline fills were possible, interpolating the previous and the subsequent visible marker positions. This was only applied if there were not more than five missing frames. In the seldom case of a larger gap, the particular trial was disregarded.

For calculating the courses of velocity and acceleration from the positional data, the numerical derivatives were derived via finite differences. A central differencing method using three points was applied.

\subsubsection*{IMU Data}
The IMU data from the Xsens sensor was saved in the proprietary Xsens format and converted to CSV files. One file was recorded per trial. Furthermore, for each session (left or right hand for each participant) the randomized delays between visual cue and acoustic start signal, the order of targets appearing, and the individual start and stop timestamps of each trial were stored.

Generally, IMU data is often used for orientation estimation. However, for the presented dataset, only the accelerometer data were used for position determination. Though, measures to compensate for sensor drift needed to be introduced. In preliminary tests, the gyroscope and magnetometer data were incorporated as well to calculate estimated corrections of the accelerations by these additionally measured values and the application of an extended Kalman filter. At the cost of computational requirements, this did not show any improvement of the acceleration data or, after integration, velocity and position estimation, respectively. For this reason, and to apply the least possible preprocessing but as much as necessary, the extended Kalman filter approach was eventually not applied. The finally conducted steps are described in the following.

First, for the acceleration data, the free acceleration was calculated by the embedded processing of the IMU sensor itself. For this, the gravitational acceleration was subtracted from the measured acceleration data. The gravitational acceleration can be measured during resting state without any external movement.

For acceleration drift compensation, movement initiation was defined as the time point when an empirically determined acceleration noise threshold combining all three dimensions of the acceleration vector in the form of Euclidean norm is exceeded. For this, the remaining noise acceleration during resting state after the end of the movement was utilized, precisely the per-component maximum of the last ten samples.

The acceleration values during initiation and ending resting state (phases under acceleration threshold) were corrected by subtracting the averaged noise activity. The averaged acceleration sensor drift rate was then calculated between movement initiation and end. With this drift rate, the intermediate acceleration measurements at each time point during movement could be corrected by the relative proportion of overall drift.

To implement the integration of acceleration to obtain velocity, an approach related to drift correction was followed, called zero velocity update (ZUPT), originally proposed for navigation and survey applications\cite{survey}. Following this method, the acceleration time course was split into continuous segments of under-or-equal-threshold and over-threshold activity, again depending on the same empirically determined acceleration threshold. For all samples, a correction of velocity due to drifting sensor data took place, while during the phases of under-threshold activity the velocity was reset to zero. This could prevent the integrative accumulation of drift errors. First, the velocity was calculated kinematically as if there was no drift: $$v_{t+1,pred} = v_t + a_t\cdot dt$$

For the first sample of an under-threshold phase, the velocity drift rate between the current and the last sample of the previous under-threshold region was calculated via dividing their velocity difference by time (sample count). Ideally, at these samples, the acceleration activity would be zero if there was no drift. Within the under-threshold phases (particularly including the very beginning of the movement, where there exists no previous under-threshold phase), the velocity was set to zero. With the drift rate, the predicted velocity values of the over-threshold activity phase laying directly in between the under-threshold phases were corrected by subtracting the corresponding drift in order to guarantee continuity.

In principle, zero acceleration could also mean constant velocity instead of zero velocity. However, this was practically not relevant due to highly variable acceleration courses and the accelerometer sensitivity to close-to-constant velocities. This led to accelerometer responses even to low deviations such as noise.

Finally, with the zero-velocity-updated integration of acceleration to obtain the course of velocity, the position course is kinematically computed based on these corrected velocity values: $$s_{t+1} = s_t + v_t\cdot dt + \frac{1}{2}\cdot a\cdot dt^2$$
\section*{Data Records}
All data can be made available upon request to the authors after publication. An analytical study of the movement data is in progress.

After extracting the data archive, the captured records can be found in the form of CSV files, structured in three main directories (the session ID \texttt{[SesID]} is composed of the subject number in the interval [1, 31], sorted by age, and the recorded hand in \{L, R\}): 
\begin{enumerate}
    \item \texttt{measurement} for the recorded data without further processing (for Vicon, the data is just split into single files and preprocessed by Nexus software): \begin{itemize}
        \item ordered list of all targets in the session:\\ \texttt{[SesID]\_targets.txt},
        \item ordered lists of recording timestamps for all trials' start and end times in the session:\\ \texttt{[SesID]\_timestampsRecStart.txt} and\\ \texttt{[SesID]\_timestampsRecStop.txt}, respectively\\ (format \texttt{YYYY-MM-DD-hh-mm-ss.sss}), as well as
        \item ordered list of the added randomized time delays for all recording trials in the session:\\ \texttt{[SesID]\_timeDelays.txt}, in milliseconds,
        \item positions for Vicon data,\\ CSV file \texttt{[SesID]\_V\_measurement\_rec[0,89]}: \begin{itemize}
            \item frame number (1 columns),
            \item \{x,y,z\} positions of marker 1 $[m]$ (3 columns),
            \item \{x,y,z\} positions of marker 2 $[m]$ (3 columns).
        \end{itemize}
        \item IMU values for Xsens data,\\ CSV file \texttt{[SesID]\_X\_measurement\_rec[0,89]}: \begin{itemize}
            \item \{x,y,z\} acceleration $[ms^{-2}]$ (3 columns),
            \item \{x,y,z\} free acceleration $[ms^{-2}]$ (3 columns),
            \item gyroscope values (angular velocities [$rad^{-1}$], 3 columns),
            \item magnetometer values (quaternion, 4 columns),
        \end{itemize}
    \end{itemize}
     \item \texttt{original} for the measurement data processed by: \begin{itemize}
        \item filtering as described above,
        \item correction for acceleration data (drift, ZUPT),
        \item time alignment as mentioned above, as well as
        \item differentiation of positional Vicon data to derive velocity and acceleration,\\ CSV file \texttt{[SesID]\_V\_original\_rec[0,89]}: \begin{itemize}
            \item time [s] (1 column),
            \item \{x,y,z\} positions $[m]$ (3 columns),
            \item \{x,y,z\} velocities $[ms^{-1}]$ (3 columns),
            \item \{x,y,z\} accelerations $[ms^{-2}]$ (3 columns),
        \end{itemize}
        \item and integration for IMU accelerometer data to calculate velocity and position,\\ CSV file \texttt{[SesID]\_X\_original\_rec[0,89]}: \begin{itemize}
            \item time [s] (1 column),
            \item \{x,y,z\} positions $[m]$ (3 columns),
            \item \{x,y,z\} velocities $[ms^{-1}]$ (3 columns),
            \item \{x,y,z\} accelerations $[ms^{-2}]$ (3 columns).
        \end{itemize}
        \end{itemize}
    \item \texttt{processed} for the final processed data, i.\,e. after applying to the original data: \begin{itemize}
        \item rotation to map the straight positional start-target-line onto the y-axis for each trial,
        \item leading to the same CSV columns as in the case of the \texttt{original} data,\\ CSV files: \texttt{[SesID]\_V\_processed\_rec[0,89]}, and\\ \texttt{[SesID]\_X\_processed\_rec[0,89]}, 
    \end{itemize}
    \item \texttt{Questionnaire} for the anonymized answers to the questions asked beforehand, including the EHI.
\end{enumerate}

Within the directories \texttt{measurement}, \texttt{original} and \texttt{processed}, there is a folder structure naming the number of the participant, the hand recorded (left/right), and the capturing system used (Xsens/Vicon), e.\,g. for subject 23: \texttt{23/L/X}, \texttt{23/L/V}, \texttt{23/R/X}, and \texttt{23/R/V}.

\subsection*{Example Data Record Structure}
As an example, \autoref{tab:subject1} illustrates the directory and file structure of the extracted dataset archive for subject number 1. Subject numbers in the interval [1, 31] can be chosen. The number of lines of the movement data files depend on the individual recordings and the particular measurement system (250\,Hz capturing rate for Vicon, 100\,Hz for Xsens).

\begin{table}[h]
\begin{tabular}{@{}llll@{}}
\toprule
Path & File & Lines & Description \\ \midrule
measurement/1/L     & 1\_L\_targets.txt     & 90      & Ordered list of target positions \\
measurement/1/L     & 1\_L\_timeDelays.txt     & 90      & Ordered list of added delays [ms] \\
measurement/1/L     & 1\_L\_timestampsRecStart.txt     & 90      & Ordered list of recording start times \\
measurement/1/L     & 1\_L\_timestampsRecStop.txt     & 90      & Ordered list of recording end times \\
measurement/1/L/V   & 1\_L\_V\_measurement\_rec[0,89].csv     & \#frames (@250Hz)    & Positions of both markers \\
measurement/1/L/X   & 1\_L\_X\_measurement\_rec[0,89].csv     & $\sim 3.5s @100Hz$      & IMU values (see above)   \\
measurement/1/R     & 1\_R\_targets.txt     & 90      & Ordered list of target positions \\
measurement/1/R     & 1\_R\_times.txt     & 90      & Ordered list of added delays [ms] \\
measurement/1/R     & 1\_R\_timestampsRecStart.txt     & 90      & Ordered list of recording start times \\
measurement/1/R     & 1\_R\_timestampsRecStop.txt     & 90      & Ordered list of recording end times \\
measurement/1/R/V   & 1\_R\_V\_measurement\_rec[0,89].csv     & \#frames (@250Hz)  & Positions of both markers  \\
measurement/1/R/X   & 1\_R\_X\_measurement\_rec[0,89].csv     & $\sim 3.5s @100Hz$      & IMU values (see above)   \\
original/1/L/V  & 1\_L\_V\_original\_rec[0,89].csv     & \#timesteps (@250Hz)      &  Pos., vel., acc. (filtered, aligned) \\
original/1/L/X  & 1\_R\_X\_original\_rec[0,89].csv     & \#timesteps (@100Hz)      &  Pos., vel., acc. (filtered, aligned)  \\
original/1/R/V  & 1\_L\_V\_original\_rec[0,89].csv     & \#timesteps (@250Hz)      &  Pos., vel., acc. (filtered, aligned)  \\
original/1/R/X  & 1\_R\_X\_original\_rec[0,89].csv     & \#timesteps (@100Hz)      &  Pos., vel., acc. (filtered, aligned)  \\
processed/1/L/V  & 1\_L\_V\_processed\_rec[0,89].csv     & \#timesteps (@250Hz) & Pos., vel., acc. (additionally rotated)  \\
processed/1/L/X  & 1\_L\_X\_processed\_rec[0,89].csv     & \#timesteps (@100Hz)      & Pos., vel., acc. (additionally rotated)   \\
processed/1/R/V  & 1\_R\_V\_processed\_rec[0,89].csv     & \#timesteps (@250Hz)      & Pos., vel., acc. (additionally rotated) \\
processed/1/R/X  & 1\_R\_V\_processed\_rec[0,89].csv     & \#timesteps (@100Hz)      & Pos., vel., acc. (additionally rotated)  \\ \bottomrule
\end{tabular}
\caption{Structured data for subject number 1}
\label{tab:subject1}
\end{table}

\subsection*{Example Trajectory of Processed Recording}

\begin{figure}[ht]
\centering
\includegraphics[width=\linewidth]{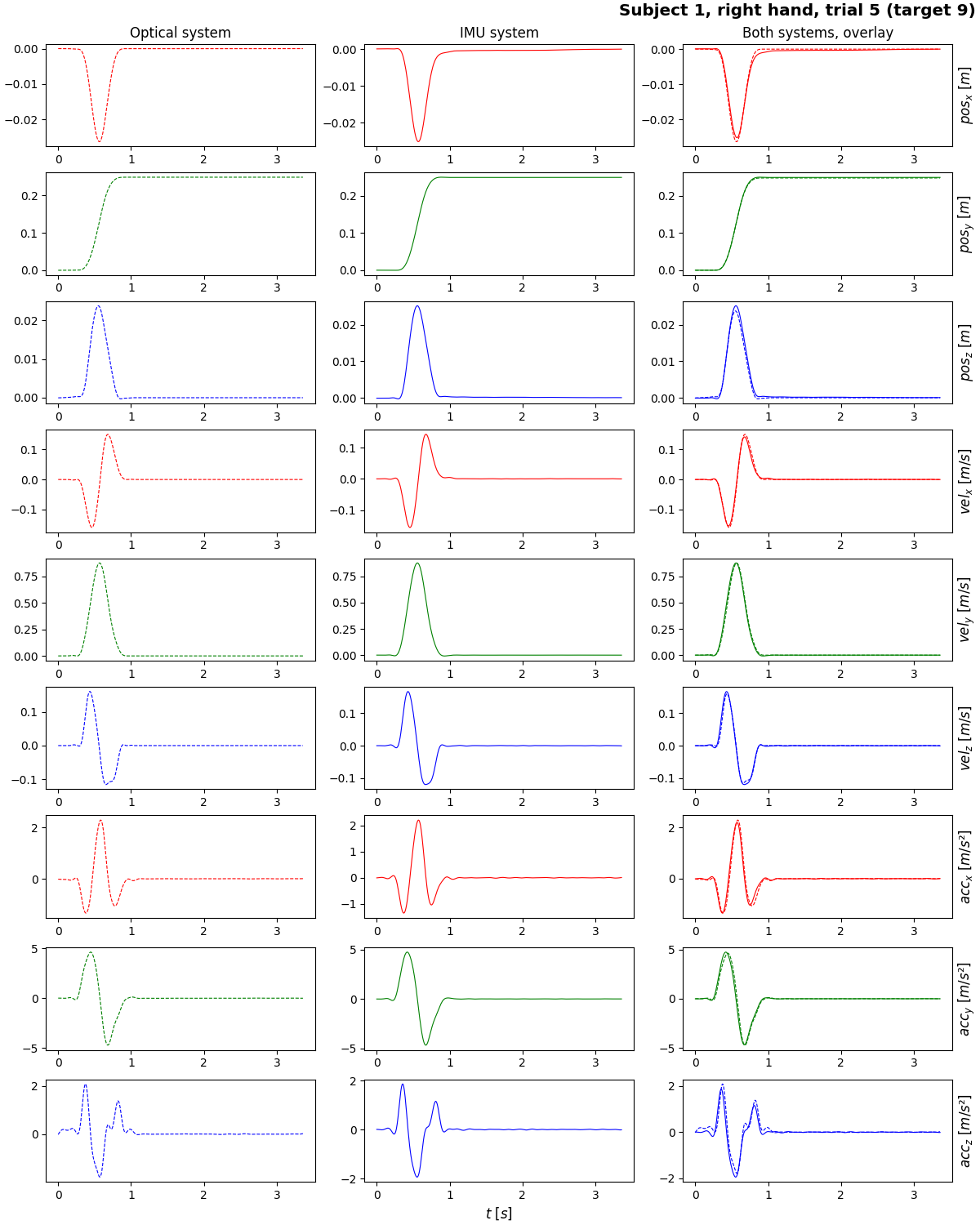}
\caption{Processed trajectory, exemplary recording of one subject, visualizing position, velocity, and acceleration for the individual dimensions over time [s]. The first two columns show the data of the individual measurement systems, the third an overlay of both systems' data.}
\label{fig:subj1}
\end{figure}

An exemplary trajectory of the processed data can be seen in \autoref{fig:subj1}, i.\,e. after filtering, drift correction, and fine alignment. This depicts one trial's position, velocity and acceleration courses over time [s] for the x-, y-, and z-dimensions of subject 1 (trial 5). While the first column shows the data from the optical system and the second column that from the IMU system, the third visualizes the overlay of both systems.

\section*{Technical Validation} \label{validation}

The Vicon Nexus motion capture system mentioned as optical reference system was calibrated with the Wand calibration method recommended by the manufacturer. 2000 calibration frames were gathered per camera, until a maximum world error of 0.17 mm per camera could be guaranteed by the software. To further reduce errors and provide redundancy, two reflection elements were attached to the cylindrical object with a fixed distance to each other. This made sure that occlusions of both markers at the same time could be avoided. When both markers were detected, the transport object's position was determined by the mean of the two markers' positions, further minimizing noise effects. In the case that for a short time, only one marker was visible to sufficiently many cameras to determine its position, the positional course of the other could be reconstructed by the software's functionality of pattern filling for single frames, orienting at the motion behavior of the visible marker.

As IMU sensors are known to be prone to disturbances, at first a pilot study was conducted in an electrically shielded lab to avoid the effect of electrostatic discharge. This analysis could not reveal any apparent difference in data quality when compared to the standard, non-shielded laboratory environment where the optical system was available. Inside the non-shielded laboratory, magnetic field mappings were conducted by means of the Xsens Magnetic Field Mapper software to find the most suitable location for the experimental setup and to calibrate the sensor with respect to the locally prevailing magnetic field. The software provided different criteria for the quality of calibration by deriving a model mapping the measured magnetic field to an ideal sphere with center at zero and a vector magnitude of 1: average of the magnetic norm close to 1, standard deviation of the norm and maximum error with respect to a norm of 1 as low as possible, no large spikes in spatial distribution of difference, and residuals of the corrected magnetic field vector following a Gaussian distribution\cite{xsens_mfm}.

The laboratory location with least temporal and spatial disturbances during the calibration process with the Magnetic Field Mapper was chosen for the experiments. Moreover, before each experimental session, further magnetic field mappings were undertaken to consider the effect of static disturbances, i.\,e. to predict and compensate for deterministic, constant magnetic field errors. As suggested by the manual from the manufacturer for 3D calibration, the IMU sensor was rotated in as many orientation angles as possible at the same location (no translational velocity) with an approximately constant angular velocity during about three minutes of magnetic field mapping.

Additionally, before each session, it was made sure that the battery of the IMU sensor was always fully charged with at least 90\% so that voltage-descent-dependent effects can be excluded as sources of errors. The wireless IMU communication with the real-time Linux kernel to avoid data loss of single frames due to process scheduling was examined in prototypical tests. These showed that all data could be successfully transferred at a rate of 100\,Hz. A fully reliable communication could be established via the Xsens wireless channel 25. In comparison to other possible channels, pilot experiments showed that this wireless channel could dependably minimize interference with other 802.11 devices (WiFi, Bluetooth).

\section*{Usage Notes}

The dataset comprises 31 subjects. The recruitment of participants by personal request does not warrant a representative selection compared to the general population.

For the motion trajectories, some trials to single targets have to be disregarded due to bad quality. In the case of the optical motion capture, this stems from occlusions that could not be fully avoided in some cases, particularly regarding both infrared reflection markers at the same time. In the case of the IMU system, the accelerometer sensor is sensitive to the exerted velocity. While fast movements, including abrupt changes of accelerations when moving the cylndrical object back onto the table, had no effect on data quality, the effect of sensor drift for slow movements could not be sufficiently compensated. However, since the paradigm was that movements had to be executed as fast and precise as possible, too slow movements appeared only in few cases. These individual trials have to be disregarded, too.

\section*{Code and Data Availability}
For recording and processing the optical motion capture data, no custom code was in use. The standard processing pipelines from Vicon Nexus 2.14 for reconstructing, marker labeling, pattern filling of single missing frames, and data export were applied.

The developed custom code for conducting the overall experiment and recording IMU data is based on an example from the Xsens MT Software Suite 2021.4 software development kit with a graphical QT user interface.

The source code for preprocessing the IMU data includes filtering, drift correction, and integration of acceleration data. In the case of the optical data, custom code for filtering and numerical differentiation of the position data was in use.

Finally, further custom code for processing both the optical- and the IMU-based data comprises synchronization by time offset adjustment and rotation for data comparability.

All custom code can be made available upon request to the authors after publication.

\bibliography{zoteroLiterature.bib,furtherLiterature.bib}

\section*{Acknowledgements}
This work was supported by the Ministry of Economics, Innovation, Digitization and Energy of the State of North Rhine-Westphalia and the European Union, grants GE-2-2-023A (REXO) and IT-2-2-023 (VAFES).

\section*{Author Contributions Statement}
\textbf{Conceptualization:} T.S., S.B., T.G., I.I.\\
\textbf{Funding acquisition:} I.I.\\
\textbf{Data recording:} T.S., S.B.\\
\textbf{Investigation:} T.S., S.B.\\
\textbf{Writing -- original draft:} T.S., S.B.\\

\section*{Competing Interests}
The authors declare no competing interests.

\end{document}